\documentclass[final]{aipproc}
\layoutstyle{6x9}
\begin{document}
\newcommand\beq{\begin{equation}}
\newcommand\eeq{\end{equation}}
\newcommand\vvt{{\bf v}_{\rm t}}
\newcommand\vvf{{\bf v}_{\rm f}}
\newcommand\vt{v_{\rm t}}
\newcommand\vf{v_{\rm f}}
\newcommand\vtt{\tilde\vt}
\newcommand\dvol{d^3\vvf}
\newcommand\Vv{{\bf V}}
\newcommand\vvtpara{{\bf v}_{\rm t ||}}
\newcommand\vtpara{v_{\rm t ||}}
\newcommand\vvtperp{{\bf v}_{\rm t\perp}}
\newcommand\vtperp{v_{\rm t ||}}
\newcommand\Vvpara{\Vv_{||}}
\newcommand\Vvperp{\Vv_{\perp}}
\newcommand\sigz{\sigma_0}
\newcommand\sigm{\sigma_{\rm m}}
\newcommand\Erf{{\rm Erf}}
\newcommand\Csi{\Xi}
\newcommand\lnlb{\ln\bar{\Lambda}}
\newcommand\lnlbb{<\ln\bar{\Lambda}>}
\newcommand\Hu{{\rm H}_1}
\newcommand\Hd{{\rm H}_2}
\newcommand\Eamu{{\rm E}_{a-1/2}}
\newcommand\Eamt{{\rm E}_{a-3/2}}
\newcommand\mm{<m>}
\newcommand\MR{{\cal R}}
\newcommand\mo{m_1}
\newcommand\md{m_2}
\newcommand\no{n_1}
\newcommand\nd{n_2}
\newcommand\mi{m_i}

\title{Dynamical Friction \\ from field particles with 
a mass spectrum}


\author{L. Ciotti}{
  address={Dept. of Astronomy, University of Bologna,\\
via Ranzani 1, 40127 Bologna, Italy}
}

\begin{abstract}

  The analytical generalization of the classical dynamical friction
  formula (derived under the assumption that all the field particles
  have the same mass) to the case in which the masses of the field
  particles are distributed with a mass spectrum is presented. Two
  extreme cases are considered: in the first, energy equipartition is
  assumed, in the second all the field particles have the same
  (Maxwellian) velocity distribution. Three different mass spectra are
  studied in detail, namely the exponential, discrete (two
  components), and power--law cases. It is found that the dynamical
  friction deceleration can be significantly stronger than in the
  equivalent classical case, with the largest differences (up to a
  factor of 10 or more in extreme cases) arising for test particle
  velocities comparable to the mass-averaged velocity dispersion of
  the field particles. The present results are relevant to our
  understanding of the dynamical evolution of globular clusters, in
  particular in the modelization of mass segregation and sedimentation
  of Blue Straggler stars and Neutron stars, and for the study of
  binary black holes in galactic nuclei.

\end{abstract}

\maketitle

\section{Introduction}

Dynamical Friction is a very interesting physical phenomenon, with
important applications in Astrophysics (and in Plasma Physics). At the
simplest level, it can be described as the slowing--down of a test
particle moving in a sea of field particles, due to the cumulative
effect of long--range interactions (no geometrical collisions are
considered). Several approaches have been devised to understand the
underlying physics (which is intriguing, as the final result is an
irreversible process produced by a time--reversible dynamics). Here I
recall the kinetic approach pioneered among others by Chandrasekhar,
Spitzer and von Neumann (e.g., see \cite{Neumann43}-\cite{Spitz87};
for a more readable mathematical account see also
\cite{Ogor65}-\cite{BT08}). More sophisticated approaches, based on a
different physical description of the phenomenon (e.g., taking also in
account the mutual interactions of the field particles, and more
realistic inhomogeneous systems), have been also developed and applied
to the case of spherical systems (e.g., see \cite{TW84} and references
therein).  A very large body of literature has been dedicated to the
study of the astrophysical consequences of dynamical friction in
astronomical systems, ranging from the sinking of globular clusters
within their host galaxy, to the formation of cD galaxies, to the
dynamical evolution of binary black holes in galactic nuclei (e.g.,
see \cite{TOS75}-\cite{AB07}).  Differences have been found between
dynamical friction in Newtonian gravity with Dark Matter and in
equivalent MOND systems (\cite{CB04,Nipoti08}); dynamical friction has
been also considered when the gravitational drag is produced by a
gaseous (instead of discrete) {\it wake} behind the test object
(e.g. \cite{Fathi10}, and references therein). An extension of the
theory to systems anisotropic in the velocity space has been also
developed (\cite{Binney77}).

In the classical approach to dynamical friction all the field
particles have the same mass, their distribution is uniform in
configuration space, and isotropic in the velocity space. Curiously,
in the enormous literature on the subject, the case of a {\it mass
  spectrum} of the field particles has not attracted much
attention. Presumably, the reason behind is the expectation that a
very massive test object, several orders of magnitude heavier than the
field masses (as often is the case in astrophysical application), should
experience the same drag force in a mass spectrum as in the classical
case, provided the total mass density of field particles is the same
in the two cases.

However, as we will see, {\it there are} astrophysical situations in
which a mass spectrum can have relevant effects, namely when the test
particle (even though very massive) travels with a velocity comparable
to the velocity dispersion of field particles, or when its mass is of
the same order of magnitude of the average mass of the field
masses. When the two features are present, the dynamical friction
evaluated in the classical case can be underestimated up to a factor
of 10 or more, with important consequences for dynamical friction
times. A specific example is represented by the population of Blue
Straggler stars (BSS) in globular clusters (e.g., see
\cite{Ferraro09}). In fact, BSS are believed to be originated by
merging or mass accretion on otherwise normal stars, so that their
mass is at most a factor of few larger than the average mass of the
stars in the parent cluster, and their mean velocities are similar to
those of the normal field stars; in addition, the stars of the
globular clusters are characterized by a mass spectrum, and finally,
globular clusters are collisional systems, with relaxation and
dynamical friction times comparable to their age. Observations also
reveal that the radial distribution of BSS in globular clusters can be
bimodal. In order to understand the possible origin of such
distribution a more accurate description of dynamical friction is
needed. Other cases of test particles (with much larger masses) moving
with a velocity similar to that of field particles is represented by
binary black holes in galactic nuclei.  These examples seem to
indicate that a study of dynamical friction in a field particle
distribution with a mass spectrum is important.

\section{The classical case}

In order to set the stage for calculations to be performed in the mass 
spectrum case, we begin with a short review of the most
important logical steps used in the derivation of dynamical friction
in the classical case.  The dynamical friction deceleration on a test
mass $M$ moving with velocity $\vvt$ in a homogeneous and isotropic
distribution (both in the configuration and in the velocity space) of
identical field particles of mass $m$ and number density $n$, is
\beq
{d\vvtpara\over dt}=-4\pi G^2nm(M+m)\lnlb{\Csi(\vt)\over\vt^3}\vvt,\quad
\vt\equiv ||\vvt ||,
\label{eq:cladf}
\eeq
where $\lnlb$ is the velocity-averaged Coulomb logarithm, the
phase-space density distribution of field masses is given by
\beq
DF=n\,g(\vf),\quad
\vf\equiv ||\vvf ||,
\eeq
and $g$ is a positive function dependent on the modulus of the 
velocity of field particles, $\vvf$. Finally, the fractional velocity 
volume function is
\beq
\Csi(\vt)=4\pi\int_0^{\vt}g(\vf)\vf^2 d\vf,
\label{eq:clacsi}
\eeq
with the normalization condition $\Csi(\infty)=1$.

In the traditional approach, eq.~(\ref{eq:cladf}) can be obtained as
follows. The basic idea is to add (vectorially) the orbital
deflections of the test particle in $n$ hypothetically independent
two--body encounters with each of the field particles. As is well known,
the total velocity change along a given unbound orbit in a generic
(escaping) force field, obeying the Newton Third Law of Dynamics, is
rigorously given by
\beq
\Delta\vvt = {\mu\over M}\Delta\Vv,\quad \mu={mM\over M+m},
\eeq
where $\Vv=\vvt-\vvf$ is the pair relative velocity.  In each
encounter under the action of the $r^{-2}$ force, the vectorial
change $\Delta\Vv$ of the relative velocity is obtained by using the
solution of the hyperbolic two--body problem.

Here, however, we obtain the change $\Delta\Vvpara$ 
in the direction parallel to the initial relative velocity by
using the impulsive approximation combined with energy conservation
along the relative orbit. For each pair it can be proved that the
change of the relative velocity {\it perpendicular} to the initial
relative velocity $\Vv$ (of modulus $V=||\Vv ||$) is
\beq
\mu ||\Delta\Vvperp||\sim {2GMm\over b V}.
\eeq
The formula above is asymptotically exact in the limit of large
impact parameter $b$ or large initial relative velocity $V$. In this
case energy conservation along each relative orbit, $V^2=||\Vv
+\Delta\Vvpara +\Delta\Vvperp ||^2$, shows that to the first order
(consistent with the adopted impulsive approximation)
\beq
\Delta\vvtpara ={\mu\Delta\Vvpara\over M}\sim
-{\mu\over M}{||\Delta\Vvperp ||^2\over 2 V^2}\Vv =
-{2G^2m(M+m)\over b^2V^4}\Vv .
\eeq
Note that the dependence of $||\Delta\vvtpara||$ as the inverse of the
cube of the initial relative velocity is asymptotically correct only
in the impulsive approximation: for slow or grazing orbits the
functional dependence of $||\Delta\vvtpara||$ on $V$ is different.
However, as in gravitational plasmas there is no screening effect, it
can be proved that the main contribution to dynamical friction comes
mainly from distant interactions (e.g. \cite{Spitz87}) so that the
above term is the leading term. In any case, it is worth to recall
that a calculation with the full solution of the two-body problem is
straightforward.

We have now to sum over all the encounters. Simple geometry shows that
their number in the time interval $\Delta t$, impact
parameter between $b$ and $b+db$, and with field particles in the
differential velocity volume $\dvol$ is
\beq
\Delta n_{\rm enc} = 2\pi b db\, ||\vvt-\vvf ||\Delta t\,n g(\vf)\dvol .
\eeq
Therefore, the differential change of the test particle velocity parallel
to the initial relative velocity is
\beq
{\Delta\vvtpara\over\Delta t} = 
-{4\pi G^2nm(M+m)g(\vf)\Vv\over bV^3}db\dvol .
\eeq
Integration over the impact parameter is a delicate step.  In fact, in
the impulsive approximation an artificial divergence appears for
$b=0$.  From the full solution of the two body problem it is easy to
show that such divergence disappears (but the divergence for
$b\to\infty$ cannot be eliminated in an infinite system).  The final
result after integration over the impact parameter can be expressed by
introducing the {\it Coulomb logarithm} $\ln\Lambda$, where the
quantity $\Lambda$ depends\footnote{Actually, the exact integration
  over the impact parameter based on hyperbolic orbits leads to the
  expression $0.5\ln(1+\Lambda^2)\simeq \ln\Lambda$, where
  $\Lambda=b_{\rm max}/[G(M+m)]$ and $b_{\rm max}$ is a fiducial
  maximum impact parameter (e.g., see \cite{BT08}).} on
$M,m,V$. Equation (8) becomes
\beq
{d\vvtpara\over dt}=-4\pi G^2nm(M+m)\ln\Lambda {g(\vf)(\vvt-\vvf)\over 
||\vvt-\vvf||^3}\dvol .
\eeq
We now integrate over the velocity space. Following Chandrasekhar
(\cite{Chandra60}), we introduce the {\it velocity weighted Coulomb
  logarithm} $\lnlb$, and therefore the Newton theorem on spherical
shells (here applied to velocity space given the assumed isotropy of
the velocity distribution of field particles), leads to the identity
\beq
\int\ln\Lambda {g(\vf)(\vvt-\vvf)\over ||\vvt-\vvf||^3}\dvol =
\lnlb{\Csi(\vt)\over\vt^3}\vvt ,
\eeq
which proves eq.~(1).  The cumulative effect of the encounters is to
slow-down the test particle in the direction of the test particle
velocity itself.  This is not trivial, as according to eq.~(6) the
deceleration in each single encounter is parallel to the {\it
  relative} velocity, and not to $\vvt$.  However, when summing over
all the encounters, the average value of the field velocity component
vanishes by assumption of isotropy.

We conclude this preparatory Section by recalling that in the commonly
considered case of a Maxwellian velocity distribution for the field
particles, the function $g$ in eq.~(2) and the velocity volume
function in eq.~(3) are
\beq
g(\vf)={e^{-\vf^2/(2\sigz^2)}\over (2\pi)^{3/2}\sigz^3},\quad
\Csi(\vt)=\Erf(\vtt)-{2\vtt e^{-\vtt^2}\over\sqrt{\pi}},
\eeq
where $\vtt\equiv\vt/(\sqrt{2}\sigz)$ is the 
normalized test particle velocity, and 
\beq
\Erf(x)={2\over\sqrt{\pi}}\int_0^xe^{-t^2}dt
\eeq
is the standard Error Function.  A final comment, of central
importance in the following discussion, is in order here.  According
to eqs.~(1) and (3) only field particles {\it slower} than the test
particle contribute to its deceleration.  This sharp ``cut'' in
velocity space results from the different assumptions, namely 1) that
the velocity distribution of field particles is isotropic, 2) that we
can take the Coulomb logarithm outside the integral in eq.~(10), and
finally 3) that the velocity change in each encounter is exactly
proportional to $V^{-2}$ (as in the first order impulsive approximation
adopted here). A more general analysis can be done, in which the
(small) correcting terms can be explicitly evaluated (e.g., see
\cite{White49}). In any case, in the presence of a mass spectrum of
field particles at equipartition, the resulting ``drag'' force is
determined by the combined effect of the mass function (in
astrophysical applications usually peaked at low masses) {\it and} the
fact that the more massive particles, responsible for large
decelerations, move slower; therefore, in principle there is an interesting
compensating effect between number density, mass of field particles
relative to the test particle, and number density in velocity space.

\section{Mass spectrum: the general case}

With the previous preparatory work, it is now easy to generalize the
classical dynamical friction formula (1) to the case of a mass
spectrum of field particles.  A generic mass spectrum with isotropic
velocity distribution is described in phase-space, by extension of the
classical treatment, with a function
\beq
DF=\Psi(m)\, g(\vf ,m),
\eeq
where 
the associated total number density of field particles and the
average mass of the spectrum $\Psi(m)$ are
\beq
n=\int_0^{\infty}\Psi(m)dm,\quad
n\mm=\int_0^{\infty}m\Psi(m)dm,
\eeq
so that 
the normalization of the velocity distribution for each mass
component leads to the condition
\beq
\Csi(\vt ,m)=4\pi\int_0^{\vt}g(\vf ,m)\vf^2 d\vf,\quad
\Csi(\infty ,m)=1\quad\forall m.
\label{eq:mscsi}
\eeq
In order to compare the dynamical friction in presence of a mass
spectrum with the classical case, we must carefully define the concept
of the {\it equivalent} classical system.  We will say that a
classical system is equivalent to a mass spectrum case if 1) the
number density in the classical case is the same as the {\it total}
number density in the mass spectrum case; 2) the field mass $m$ in the
classical case is the same as the {\it average} field mass $\mm$; 3)
the velocity dispersion of the Maxwellian velocity distribution in the
classical case is the same as the {\it equipartition} velocity
dispersion of the mass spectrum case.  We can summarize the above
conditions by saying that the comparison is between two systems with
the same number, mass, and kinetic energy density of the field
particles. Similar comments, but different answers, apply when instead
of equipartition among the different species, all the field particles
with a mass spectrum share the same velocity distributions (for example
as expected in a collisionless system made of stars and dark matter).

In the equipartition case, we use as 1--dimensional equipartition
velocity dispersion the one relative to the average mass, i.e., we assume
\beq
m\sigm^2=\mm \sigz^2,\quad
g(\vf ,m)={e^{-\vf^2/(2\sigm^2)}\over (2\pi)^{3/2}\sigm^3}=
{e^{-r\vf^2/(2\sigz^2)}r^{3/2}\over (2\pi)^{3/2}\sigz^3},\quad
r\equiv{m\over\mm},
\eeq
so that from eq.~(\ref{eq:mscsi}) 
\beq
\Csi(\vt ,m)=\Erf(\vtt\sqrt{r})-
{2\vtt\sqrt{r}e^{-\vtt^2 r}\over\sqrt{\pi}},
\label{eq:csimaxwm}
\eeq
where again $\vtt =\vt/(\sqrt{2}\sigz)$.
In the present case the differential number of encounters suffered by 
the test particle is 
\beq
\Delta n_{\rm enc} = 2\pi b db\, ||\vvt-\vvf ||\Delta t 
\, \Psi (m) g(\vf ,m)dm\dvol .
\eeq
Therefore, by summing the formula obtained in the
classical treatment over all the species, the deceleration in the mass
spectrum case is given by
\begin{eqnarray}
{d\vvtpara\over dt}&=&
-4\pi G^2\lnlbb{\vvt\over\vt^3}\int_0^{\infty}\Psi(m) m(M+m)\Csi(\vt ,m)dm\cr
&=& -4\pi G^2n\mm (M+\mm)\lnlbb {\Csi^*(\vt)\over\vt^3}\vvt ,
\label{eq:msdf}
\end{eqnarray}
where now $\lnlbb$ is the mass-averaged Coulomb logarithm.  The second
of the above equations is just the definition of the new velocity
coefficient $\Csi^*$. In practice, from the knowledge of this last
function one can derive the dynamical friction deceleration in case of
a mass spectrum by using the same formalism of the classical case,
where $m$ is replaced by $\mm$.  In all the following computations we
will assume that $\lnlbb\simeq\lnlb$.

It is important to note that, for {\it large} velocities of the test
particle the velocity volume factor $\Csi(\vt,m)$ tends to unity, and
therefore, {\it the values of the dynamical friction deceleration in
  the high--velocity limit can be also interpreted as the scaling
  factor between the classical and mass spectrum case when all the
  species in the mass spectrum have the same velocity
  dispersion}. This case is of astrophysical importance, for instance
for dark matter halos in galaxies, where dark matter particles and
stars likely are {\it not} at the equipartition.  We now study a few
explicit cases of mass spectrum amenable to analytic solutions, so
that the differences with the equivalent classical cases can be
quantified.

\section{Exponential spectrum}

In this case the mass spectrum is given by 
\beq 
\Psi(m)={n e^{-m/\mm}\over\mm}.  
\eeq 
The integral over masses in eq.~(\ref{eq:msdf}) can be
performed analytically by inverting the order of integration between $m$
and $\vf$. The result is 
\beq 
\int_0^{\infty} \Psi(m)
m(M+m)\Csi(\vt ,m)dm =n\mm^2[\MR\Hu(\vtt)+\Hd(\vtt)],\quad
\MR\equiv{M\over\mm}. 
\label{eq:csiexp}
\eeq 
In practice, the mass ratio $\MR$ measures the mass of the test
particle in units of the average mass of the field particles. 
From eqs.~(\ref{eq:msdf}) and (\ref{eq:csiexp}) it follows that the 
associated velocity factor can be written as
\beq
\Csi^*(\vt)={\MR\Hu(\vtt)+\Hd(\vtt)\over\MR +1},
\eeq
with the surprisingly simple result
\beq
\Hu(\vtt)={\vtt^3(5+2\vtt^2)\over 2(1+\vtt^2)^{5/2}},\quad
\Hd(\vtt)={\vtt^3(35+28\vtt^2+8\vtt^4)\over 4(1+\vtt^2)^{7/2}}.
\label{eq:funexp}
\eeq
\begin{figure}
  \includegraphics[height=.5\textheight]{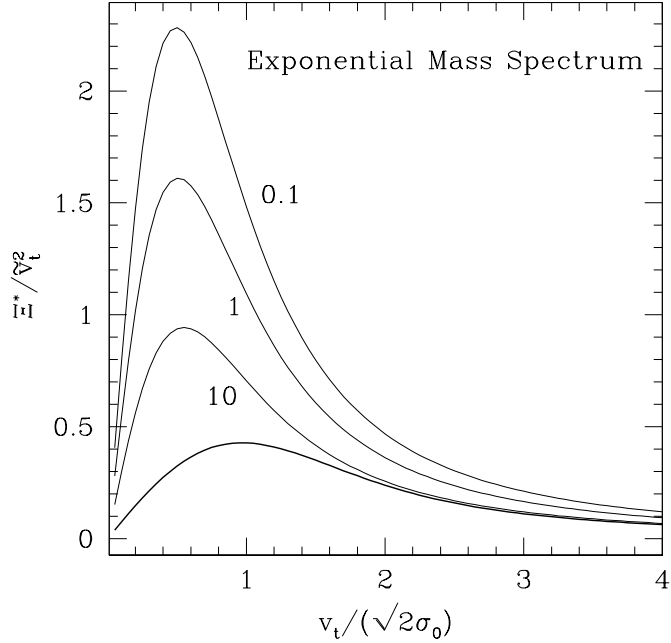}
  \caption{The velocity coefficient $\Csi^*/\vtt^2$ in eq.~(19) for
    the Exponential Mass Spectrum case with equipartition;
    $\vtt=\vt/(\sqrt{2}\sigz)$.  The curves (from top to bottom)
    correspond to a test particle with mass 0.1, 1, and 10 times the
    average mass of the spectrum, respectively. Mass ratios larger
    than $\sim 10$ produce curves almost identical to the $\MR=10$
    case. The heavy solid line represents the velocity coefficient of
    the equivalent classical case, i.e.  when the field masses are all
    identical, and their number density, average mass, and kinetic
    energy density are the same as in the mass spectrum case.}
\end{figure}
As expected, eqs.~(22) and (23) prove that the result coincides
asymptotically with the classical case for fast ($\Hu\sim 1$ and
$\Hd\sim 2$) {\it and} massive ($\MR\gg 1$) test particles. In
general, as can be seen from Fig.~1, the velocity factor in the case
of exponential mass spectrum with equipartition is larger than in the
corresponding classical case (heavy line): for massive test particles
the maximum drag (corresponding to $\vt\simeq 0.81\sigz$) is a factor
$\approx 2$ higher than in the equivalent classical case.  The
dynamical friction time is correspondingly shorter, with significant
discrepancies for test particles moving with velocities comparable to
the equipartition velocity dispersion of the field particles. Finally,
the leading term of eq.~(22) for $\vt\to\infty$ shows that in the non
equipartition case the correcting factor to be adopted when using the
classical formula is $(2+\MR)/(1+\MR)$, so that for $\MR$ of order of
unity the classical formula underestimates the dynamical friction
deceleration by a factor $\approx 1.5$.

\section{Discrete spectrum}

For the case of a system made of two species of field particles the mass 
spectrum is 
\beq 
\Psi(m)=\no\delta(m-\mo)+\nd\delta(m-\md).
\label{eq:psids}
\eeq 
With the convenient introduction of the dimensionless parameters
$x\equiv \nd/\no$ and $y\equiv\md /\mo$, it follows that
\beq 
n=(1+x)\no ,\quad \mm={\no\mo +\nd\md\over \no+\nd}={1+xy\over 1+x}\mo;
\eeq 
the limit $y=1$ recovers the classical case. The generalization to an
arbitrary number of different field components presents no
difficulties.  The mass integration in eq.~(\ref{eq:msdf}) is
immediate, the result is formally identical to eqs.~(21) and (22), while 
from eqs.~(24) and (17) we now have
\beq
\Hu={\Csi(\vt,\mo)+xy\Csi(\vt,\md)\over 1+xy},\quad
\Hd={(1+x)[\Csi(\vt,\mo)+xy^2\Csi(\vt,\md)]\over (1+xy)^2}.
\eeq
For low velocities of the test particle one finds the
asymptotic trends
\beq
\Hu\sim{4\vt^3 (1+x)^{3/2}(1+xy^{5/2})\over 3\sqrt{\pi}(1+xy)^{5/2}},
\quad
\Hd\sim {4\vt^3 (1+x)^{5/2}(1+xy^{7/2})\over 3\sqrt{\pi}(1+xy)^{7/2}}.
\eeq
In turn, for large velocities of the test particle the leading terms 
are
\beq
\Hu\sim 1,\quad
\Hd\sim {(1+x)(1+xy^2)\over (1+xy)^2}.
\eeq
Therefore, for fast and massive test particles, the dynamical friction
force in the presence of equipartition is the same as in the
equivalent classical case. In the non equipartition case, the
correcting factor for the classical dynamical friction formula is
obtained by evaluating eq.~(22) with the expansions given in eq.~(28).
\begin{figure}
\includegraphics[height=.25\textheight]{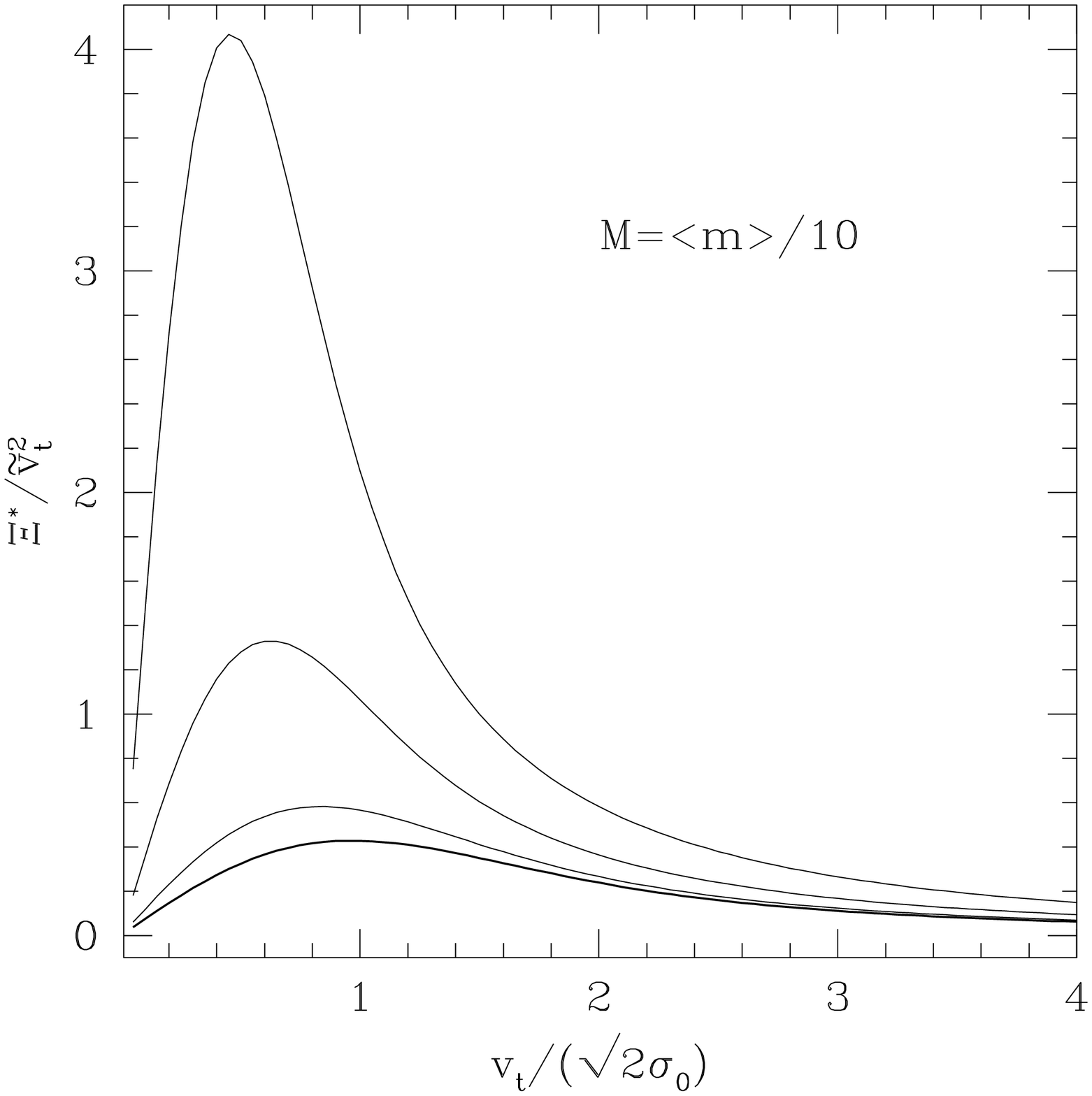}
\includegraphics[height=.25\textheight]{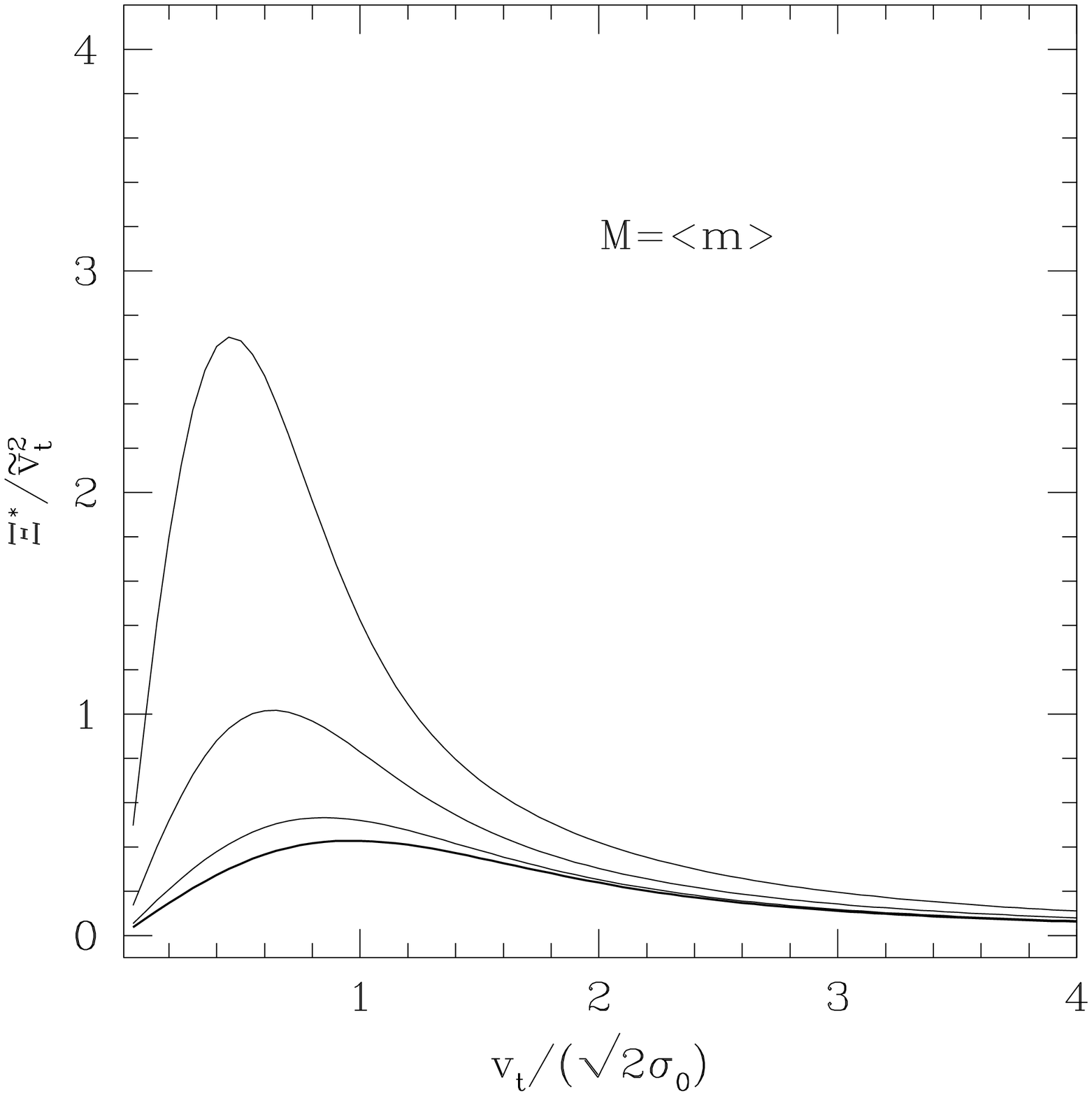}
\includegraphics[height=.25\textheight]{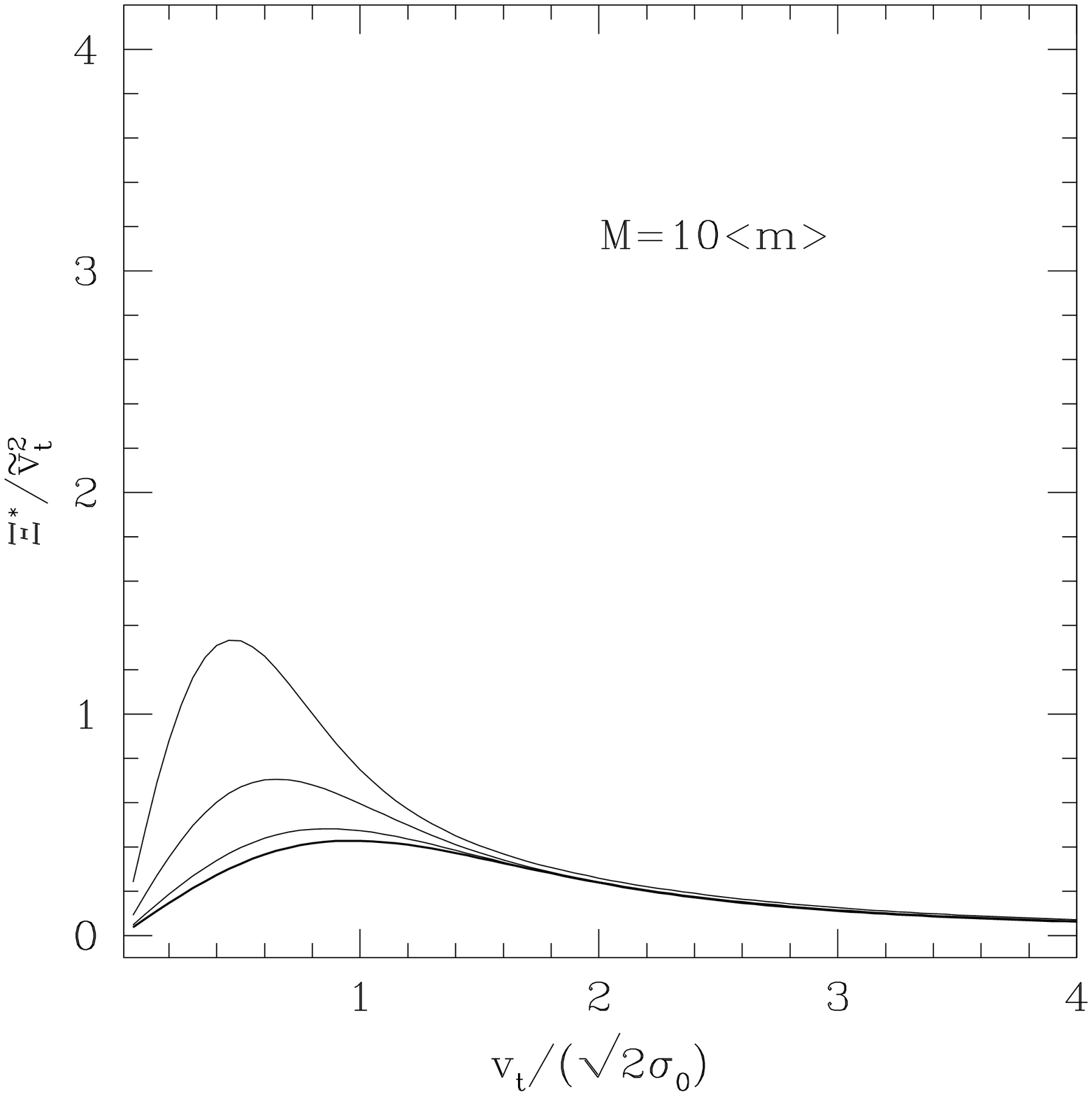}
\caption{The velocity coefficient $\Csi^*/\vtt^2$ in eq.~(19) for the
  Discrete Mass Spectrum case with two species in equipartition and
  with the same mass density, i.e. $\no\mo=\nd\md$.  The panels (from
  left to right) correspond to mass ratios $\MR =0.1, 1, 10$, while
  the curves in each panel (in decreasing order) correspond to number
  ratios $\nd/\no =x=8,4,2$, respectively. For increasing mass
  ratio $\MR$ and large velocities of the test particle the
  deceleration tends to the value obtained in the equivalent
  classical case.}
\end{figure}

We now study the case of arbitrary mass ratios and velocities. For
simplicity we restrict the following analysis to the special case of a
system in which the densities of the species 1 and 2 are the same,
i.e., $\no\mo=\nd\md$. From the definitions in eq.~(25) it follows
that $xy=1$. Therefore, for $x>1$ the masses $\md$ are lighter and
more numerous than the species 1; it is easy to recognize that the
cases $x>1$ and $x<1$ (with reciprocal values) are coincide. In Fig.~2
the situation is illustrated for three different mass ratios $\MR$ and
different number ratios of the two field species. The qualitative
trend is the same as in the exponential case: the equivalent classical
case always underestimates the true value of dynamical friction, with
largest deviations (at fixed $\MR$) for test particle velocities
comparable to the field equipartition velocity dispersion. The
discrepancies can be as large as a factor 6 - 10 for masses of the
test particle of the same order of magnitude of the average mass of
the spectrum.  Similar calculations can be done on the other relevant
case of identical {\it number} density of the two species, $\no=\nd$
(i.e., $x=1$), and again the results for the mass spectrum case shows
that the frictional force is stronger than in the equivalent classical
case.

\section{Power--law spectrum}

As commonly done in many cases of astrophysical interest, 
we finally assume a power--law
spectrum peaked at low masses, with a minimum mass $\mi$, a finite
average mass $\mm$, and exponent $a >1$, i.e.
\beq 
\Psi(m)={na\mi\over m^{1+a}},\quad 
\mm={a\mi\over a-1},\quad
m\geq\mi .  
\eeq 
As in the two previous cases, mass integration in eq.~(19) can be done 
analytically. Equations (21) and (22) remain unchanged, while now
\beq
\Hu(\vtt)=\Erf(\vtt\sqrt{c})-
{(2a-3)\sqrt{c}\vtt\Eamu(c\vtt^2)\over\sqrt{\pi}},
\eeq
\beq
\Hd(\vtt)={(a-1)^2\over a(a-2)}\left[
\Erf(\vtt\sqrt{c})-
{(2a-5)\sqrt{c}\vtt\Eamt(c\vtt^2)\over\sqrt{\pi}}\right].
\eeq
The convergence of the $\Hd$ function in integral (21) requires $a>2$.
This condition may be relaxed if a cut--off on large
masses is applied to the mass spectrum.  The transcendent function
appearing in the two expressions above is the Exponential Integral,
which is related to the left incomplete Euler Gamma Function as
\beq
{\rm E}_k(z)\equiv\int_1^{\infty}
t^{-k}e^{-tz}dt=z^{k-1}\Gamma(1-k,z).
\eeq
It is easy to show that, for large velocity of the test mass, 
asymptotically
\beq
\Hu\sim 1,\quad
\Hd\sim {(a-1)^2\over a(a-2)}.
\eeq
Therefore, this demonstrates again that for high velocity and large
mass of the test particle the classical result is recovered.  The
expansion for vanishingly small $\vt$ requires that different cases
must be distinguished.  In general, when $a>7/2$ both the $\Hu$ and
$\Hd$ functions both vanish as $\vtt^3$, while $\Hu={\cal O}(\vtt^{2a-2})$
for $a<5/2$ and $\Hd={\cal O}(\vtt^{2a-4})$ for $a<7/2$. In the
critical cases the functions $\Hu$ and $\Hd$ vanish as ${\cal
  O}(-\vtt^3\ln\vtt)$.  Note that the function $\Csi^*/\vtt^2$
diverges for $\vtt\to 0$ when $a<3$ as a consequence of the $\Hd$
behavior, while it reaches a finite value when $a=3$. In Fig.~3 some
representative case is illustrated, for different values of $\MR$ and
of the power--law index $a$. The corrective factor for the classical formula
when the field particles are not at the equipartition, but are characterized 
by the same Maxwell distribution independently of their mass, is obtained by 
inserting the functions in eq.~(33) in eq.~(22).

\begin{figure}
\includegraphics[height=.25\textheight]{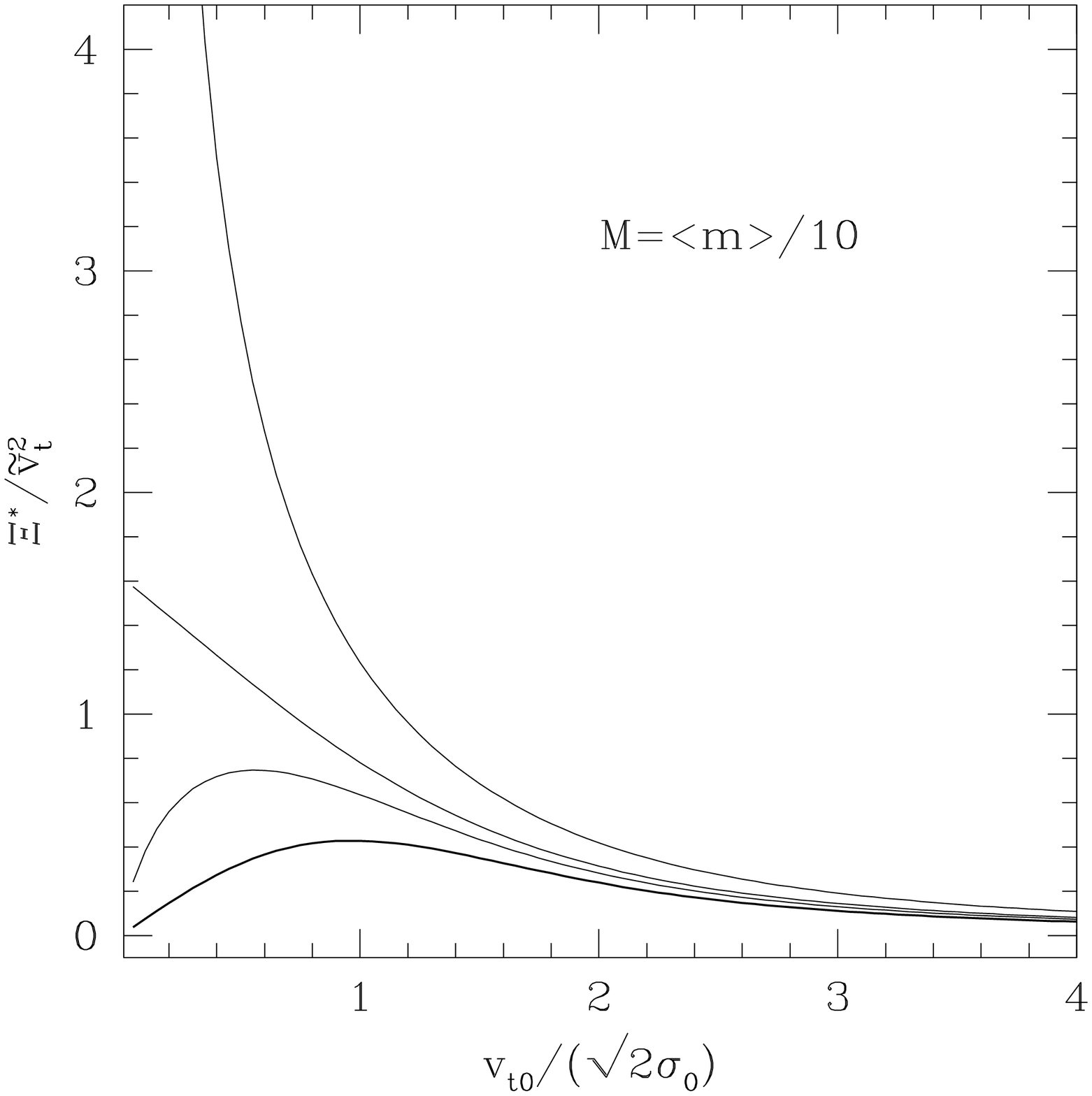}
\includegraphics[height=.25\textheight]{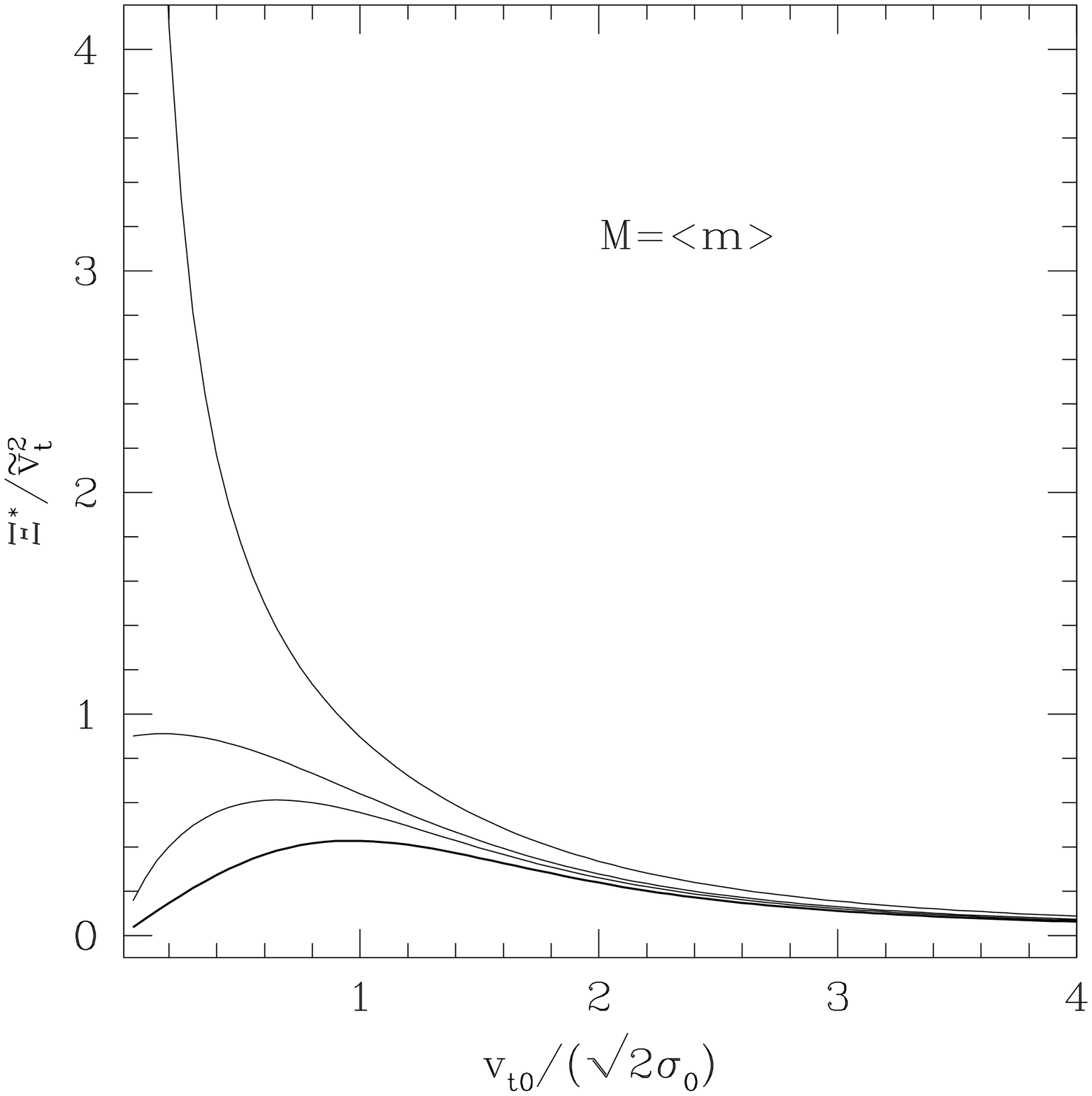}
\includegraphics[height=.25\textheight]{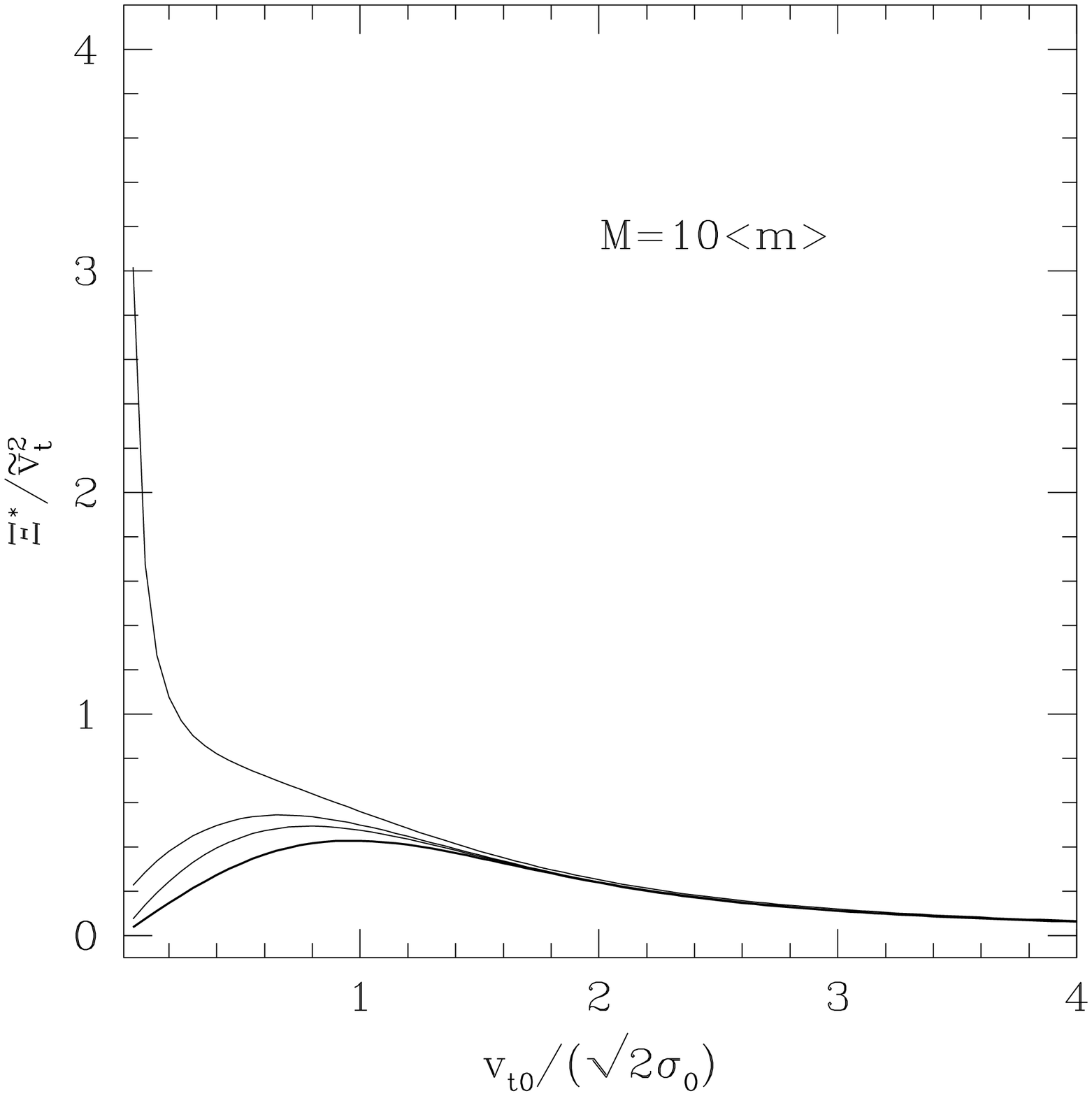}
\caption{The velocity coefficient $\Csi^*/\vtt^2$ in eq.~(19) for the
  Power-Law Mass Spectrum case in equipartition.  The panels (from
  left to right) correspond to mass ratios $\MR = 0.1, 1, 10$, while
  the curves in each panel (in decreasing order) correspond to
  exponent $a=2.5,3,3.5$, respectively. As in the previous cases, for
  increasing mass ratio $\MR$ and large velocities of the test
  particle the deceleration converges to the value obtained in the
  corresponding classical case.}
\end{figure}

\section{Conclusions}

In this paper I presented a generalization of the standard dynamical
friction formula to the case of a test particle moving in a
homogeneous distribution of field particles caracterized by a mass
spectrum.  Suprisingly, the mass spectrum problem has not received
much attention in the astrophysical literature.  In the present
investigation the velocity distribution of each species of the field
particles is Maxwellian, and equipartition among the species is
assumed. It has been also shown how the situation in which all the
field particles have the same velocity distribution can be easily
recovered as a limit case of the equipartition analysis.

The comparison with the classical case is done by considering an
equivalent classical system in which 1) the field particles have the
same mass as the average mass of the mass spectrum case, 2) the number
density is equal to the total number density in the mass spectrum
case, and finally 3) the velocity dispersion of the classical case
equals the equipartition velocity dispersion in the mass spectrum
case. In practice, the classical and the mass spectrum cases have the
same number density, mass density, and kinetic energy content of the
field masses. Three specific cases of mass spectrum (i.e., an
exponential mass spectrum, a two--component discrete spectrum, and a
power--law spectrum) have been considered, and the associated
analytical formulae derived.

A few common trends are noted. First, for {\it fast} and {\it massive}
test particles the results in the classical and in mass spectrum cases
are asymptotically identical, because for high velocities the velocity
volume factor tends to unity, and for large test masses the specific
form of the mass spectrum becomes irrelevant, as all the field
particles can be considered vanishingly small, and only the mass
density of field particles appears in the relevant expressions of the
friction coefficient.

Second, in {\it all} the cases considered, the dynamical friction
force in the mass spectrum case is larger than in the corresponding
classical case. The largest differences are found for test particle
masses comparable to the average mass of the spectrum, and test
particle velocities close to the equipartition velocity
dispersion. The differences can be as high as a factor of 10 or
more. The dynamical friction times are correspondingly reduced.

Third, for very large velocities of the test particle, but for a test
mass particle comparable to the average mass of the spectrum (say $\MR
< 10$), there are differences between the mass spectrum case and the
classical case. From the astrophysical point of view this last result
also applies to the case in which the mass spectrum particles are {\it
  not} at the equipartition, but the species are characterized by the
same velocity dispersion (for example, stars and dark matter particles
in the common pontential well).

It follows that the classical dynamical friction formula for a very
massive object (such as a globular cluster or a mini dark matter halo)
sinking into a larger system, made of stars and dark matter, should
provide correct values for the dynamical friction force (as far as the
sinking velocity is large).  However, there are astronomical systems
where the present investigation is relevant, i.e., the case of Blue
Straggler stars in globular clusters. In fact, 1) BSS stars have a
mass slightly larger than the average mass of the field stars in the
host system; 2) the velocity of BSS is close to the local velocity
dispersion of the field stars, just because they are orbiting in the
parent globular cluster; 3) the field stars in a globular cluster are
characterized by a mass spectrum, and the assumption of equipartition
is reasonable, because of the quasi--relaxed state of globular
clusters. If the three points above apply, then it follows that the
adoption of the classical dynamical friction formula to study the
evolution of the spatial distribution of BSS (or neutron stars) in
globular clusters may be inaccurate, with prediction of excessive
sinking times. It would be very interesting to study whether the
formulae derived in this paper succedes in explaining the observed
radially bimodal distribution of BSS in some well studied globular
cluster; an important issue here is how the initial mass function of
the field stars is modified at each radius by dynamical evaporation of
low mass stars, with the obvious consequence of a reduction of the
spanned mass interval.  Another case of possible interest is
represented by the initial stages of the dynamical evolution of binary
black holes in galactic nuclei. Finally, from a theoretical point of
view, it would also be interesting to extend the present treatment to
the evaluation of the two--body relaxation time in the presence of a
mass spectrum.

\begin{theacknowledgments}
  I whish to thank G. Bertin, J. Binney, F. Pegoraro and M. Stiavelli
  for very useful comments.
\end{theacknowledgments}

\end{document}